\documentclass[10pt,journal,compsoc]{IEEEtran}

\ifCLASSOPTIONcompsoc
  \usepackage[nocompress]{cite}
\else
  \usepackage{cite}
\fi
\usepackage{amsmath}
\usepackage{algorithmic}
\usepackage{array}
\usepackage{float}
\usepackage{graphicx}
\begin{document}

\title{Authentication and Hand-Over\\ Algorithms for IoT Groups}

\author {Y\"{u}cel~Ayd{\i}n,
       ~G\"{u}ne\c{s}~Karabulut~Kurt~\IEEEmembership{Senior Member} and {Enver~Ozdemir~\IEEEmembership{Member} }  

}

\IEEEtitleabstractindextext{
\begin{abstract}
Current advancements in mobility of devices and also Internet of Things (IoT) have replaced the central networks by distributed infrastructure. The more a network is distributed, the more the security of infrastructure and the communication is getting complex.
The members in a distributed network create different groups according to their coverage area or their requirements. Mobility nature of the members brings a problem called hand-over of members between groups. Current authentication methods are not applicable
due to the lack of resources in the devices.A lightweight authentication method and an easy and fast hand-over process are the current need for the distributed networks. 
Shamir Secret Sharing algorithm is used for the authentication process in the studies before, but still secure group authentication algorithm and hand-over process are challenges in the group authentication. In this study, a new method is proposed to provide a secure group authentication and hand-over process between groups based on Lagrange's Interpolation.
\end{abstract}

}

\maketitle
\IEEEdisplaynontitleabstractindextext
\IEEEpeerreviewmaketitle

\IEEEraisesectionheading{\section{Introduction}\label{sec:introduction}}

\IEEEPARstart{C}{ommunication} turns into a distributed structure progressively. Many devices within a certain area can exchange data intensely or new devices can get involved to the communication.
The worst case is that all the devices in the communication can be mobile. The number of mobile devices and the distributed networks will increase in the future.
\\
\\
Unmanned Aerial Vehicles(UAV)-assisted 5G cellular infrastructure for disaster-resilient networks which was proposed in [1] is an example for the distributed nature of future networks.
While some users receive service from base stations (BS) in the infrastructure, UAVs provide service for users who lost their connectivity due to the disasters.
There are four different groups in the UAV-assisted 5G cellular infrastructure: users served by UAVs, users served by mmWave BS (lower frequency),users served by mmWave BS (higher frequency),user served by $\mu$mmWave BS.
The members of these four groups will interact with BSs and each other, but a secure authentication mechanism for such a distributed infrastructure is still a challenge. 
\\
\\
The developments on the Radio Frequency Identification (RFID) and sensor network technologies will create a more and more distributed environment in the future. The traditional mobile network which uses only smart phones currently will be replaced by the heterogous network after increasing the number of sensors around us. The evolution in the network science  will create more mobile and more distributed environment[2].
\\
\\
Security is one of the challenges in the mobile, distributed and crowded networks. In some networks, it is possible to see thousands of nodes. Therefore; the identity of each node is critical for the security of whole network. Also, the confidentiality, integrity and availability of communications between nodes should be well secured. Encryption is the first option for the security of communication. There are two encryption mechanisms which are symmetric and asymmetric key encryptions. Asymmetric key encryption is time and resource consuming method for very distributed networks. Therefore; it is not the first option for IoT or UAV kind networks. The problem in the symmetric key encryption is the key management and exchange methods. It is hard to share the same key with all nodes and to update periodically.  Due to these reasons; a new lightweight encryption method is needed for very distributed networks. Shamir Secret Sharing method is a solution for such a challenging problem. It is possible to create a shared key between n users at the same time. Only the nodes that have k value can calculate the secret key. [3]
\\
\\
Authentication is the one of the most important security step for such a distributed environment. Traditional authentication process includes one claimer that requests authentication and one prover that approves the claims.
This kind of authentication process can be named as one-to-one authentication. One-to-one authentication is no longer applicable for such a kind of environment. If n users want to authenticate each other, one user should repeat the authentication process n-1 times.
O(n) is the complexity of this kind of authentication.
\\
\\
Many-to-many authentication which is named Group Authentication is the new schema for the complex, mobile and crowded networks. The main idea of Group Authentication is to authenticate n users at the same time. 
The new complexity of authentication process will be O(1).
\\
\\
There are several studies which are mentioned in the related works part of this study.The authors tried to find a way in the studies in order to authenticate users who belong to the same group at the same time. 
But current mobility of the users is extremely high and it will be more in the near future. Therefore; one user who belongs to a group will travel to the area of other groups and will try to establish communication with other groups.
Hand-over of users between different authentication groups is still a dilemma for group authentication studies. In this study, a novel approach is proposed for handing-over process in group authentication.
\\
\\
This work is organized as follows. The following section provides an overview of related works about group authentication and hand-over. In the third session, the proposal method for group authentication and authentication between two nodes from different group is denoted. The study is completed by conclusion part.

\section{Related Work}
Authors proposed a group authentication and key agreement protocol for Long Term Evolution (LTE) networks in [4]. They mentioned three different components in the proposed protocol which are Mobile Equipment (ME), Serving Network (SN) and Home Network (HN).
When any ME wants to get involved to a group, it sends request to the SN for authentication. SN sends the identity of ME to the HN. HN is responsible for identifying the ME. This process is repeated for each ME in the group.
The protocol is one-to-one authentication process and it is not applicable for very distributed networks due to the time and resource limitations. Moreover, when one mobile node wants to communicate with other group, the authentcation process should be repeated for
other group.
\\
\\
Another group authentication schema is proposed in [5]. They used a hash function with a pre-shared key (HMAC) in order to authenticate nodes. At the second phase of the authentication process, each user sends a reply to the authentication point at different times.
Second process makes the protocol one-to-one authentication schema.
\\
\\
A novel method for handover problem for WIMAX (Worldwide Interoperability for Microwave Access) Networks is proposed in [6].  In the WIMAX architecture, there are one Base Station (BS) and several Mobile Stations (MSs) in a group. Also there is connection between BS and AAA (Authentication, Authorization and Accounting) server. They used Extensible Authentication Protocol (EAP) to establish secure connection between MS and AAA server. MS and AAA server identify a Master Session Key (MSK) for further authentication after EAP connection. AAA server shares the MSK with BS after a keyed hashing operation. Then BS and MS can authenticate each other by the result of hashing. At the same time MS and BS choose random numbers in order to use in Elliptic Curve Key Exchange method. At the end of the process each side computes the secret key (SK) for further communications. 
For handover process BS shares the hashing result of SK with other BS. When MS wants to begin the authentication process, it computes the hash of SK and sends the $BS_{2}$. $BS_{2}$ confirmed the result by using the credentials that is received from $BS_{1}$. Also MS and $BS_{2}$ choose random numbers for Elliptic Curve Key Exchange Method in order to identify a secret key for further communications.
In the study, each MS needs to repeat the authentication process with $BS_{1}$ to have a group authentication. But this kind of authentication takes too much time and resource for distributed networks. Also, there is no proposal for authentication between Mobile Stations connected to the different Base Stations.
\\
\\
HashHand[7] is another proposal to hand over nodes between access points in mobile networks. The proposed structure includes the Authorization server (AS), Access point (AP) and Mobile node (MN). The AS is responsible for transactions with a high computational load.
AS chooses a secret (s) and determines G (cyclic additive group), $G_{T}$ (cyclic multiplicative group), e: G x G $\to$ $G_{T}$ (bilinear map), $P_{pub}$ (public key), $H_{1}$, $H_{2}$(Hash Functions). AS shares these public parameters except secret key.
After determining identities (IDs) for each AP, AS calculates public keys ($H_{1}$($ID_{AP}$)) and private keys (s.$H_{1}$($ID_{AP}$)). AS shares public and private keys with each AP.  AS determines a group of random numbers ($a_{1}$,$a_{1}$,...,$a_{j}$) 
and calculates public keys ($H_{1}$($a_{i}$)) and private keys (s.$H_{1}$($a_{i}$)) for each MN. After calculating of key pairs, AS share these values with relevant MNs. When any MN wants to move from $AP_{1}$ to $AP_{2}$, MN chooses random private and public key pairs. MN sends to $AP_{2}$ an authentication code. The code is computed by using private key of MN and public key of $AP_{2}$ ($H_{2}$(e($MN_{private}$,$AP_{public}$))). Once new AP gets the authentication code, AP computes confirmation code ($H_{2}$(e($MN_{public}$,$AP_{private}$))). If the codes are the same, authentication is done between MN and new AP. The proposal is a good example of implementation of Eliptic Curve Criptography (ECC) for handover purposes. Mobile nodes only consumes source in order to calculate bilinear pairing(e) for authentication code. The most source consuming jobs are done by the AS. Therefore; we can assume the proposal an one to many authentication method.
\\
\\
ECC with RSA algorithm is used in [8] in order to overcome with the vulnerabilities in HandHash. AS determines prime numbers (p,q,r,s) and computes the $\phi$(n)=(p-1)*(q-1), $\phi$(m)=(r-1)*(s-1) and $\phi$(N)=$\phi(n)$*$\phi(m)$. Then AS chooses an integer e which satisfy (1$<$e$<$$\phi$(N)) and gcd (e,$\phi$(N))=1 and map the number to the Group(G) as public key ($MN_{e}$) for each MN. Private key($MN_{d}$)  is calculated by $MN_{e}^{-1}$. Also AS makes the same operations for each AP and determines public key($AP_{e}$)  and private key($AP_{d}$)  pairs. AS share these keys and a M point in the group with relevant APs and MNs. When any MN wants to authenticate itself with any AP, MN chooses a random number (k) and computes C=k.B(B is a generator of G) and Authentication Code=M+$AP_{e}$+C. Once AP gets the code and C, it computes verification code=Auth+$AP_{d}$-C. If M and verification code is same, authentication is done between AP and MN. The algorithm works faster than HashHand and use less computational power. 
\\
\\
Harn proposed an algorithm for Group Authentication in [9]. The algorithm is built based on the Shamir's Secret Sharing Scheme. 
The authentication is not one-to-one type authentication as currently used authentication methods. 
The algorithm provides authentication for several IoT nodes at the same time. This is called many-to-many authentication type. 
\\
\\
One of the nodes selects a random polynomical $f(x)$ of degree $t-1:f(x)=a_{0}+a_{1}x+...+a_{t-1}x^{t-1} mod p$. 
The secret for the communication is $a_{0}$ value of the polynomical. The node calculates one secret and one private key for each nodes in the group. 
Then, the node distributes the keys to the nodes in the group. Each group calculates the secret by Lagrange Interpolating Formula.
 In the algorithm, many-to-many authentciation is done.However; There is no proposal for authentication of two nodes from two different groups.
\\
\\
The authors proposed an algorihtm by using Paillier Threshold Cryptography in [10]. 
They compared their result with Harn Group Authentication Method and give the results from their experiments.
\\
\\
In the proposed method, one of the nodes selects a public key and generates several private keys for each member in the group. 
After encrypting of a secret key by public key, the node distributes the encrypted secret and private keys to the corresponding nodes in the group.
 Each node decrypts the secret by its own private key and generates a Partial Decrypted Message (PDM). 
The nodes shares their PDMs with the other nodes. 
After combining all the PDMs, each node in the group gets the secret for next communications. 
\\
\\
The results from [10] shows that their algorithm has better computational time than the Harn Group Authentication Algorithm. But they didn't take into account the computational cost of public and private key encryptions.
 However, they also did not propose any method for authentication of two nodes from two different groups. 
\\
\\
Paillier Threshold Cryptography method is used in [11] in order to authenticate many devices at once. A group manager is responsible to generate a public key and a private key for each group member. Group Manager shares the partial private keys with relevant members in the preparation phase of the protocol. It is not specified in the article how to distribute private keys securely. Group Manager sends a challenge encrypted by public key and the hash of the challenge to each group member in the authentication phase. When each member receives the challenge they decrypt the challenge with their partial keys and send decrypted challenge to other members including group manager. When each member receives all decrypted challenges they combine all challenges in order to have session key. After calculating the session key each member computes the hash of the session key and compare the result with the hash that the group manager sent.
\\
\\
Chien[12] showed that the Harn schemas allow some attacks. If an attacker can get k distinct values in k different trials, the secret function choısen by Group Manager can be solved and all secrets in the algorithm can be obtained. Chien proposed a new method based on Shamir Secret Sharing, ECC and pairing-based cryptography in order to ensure a secure group authentication process. According to proposal, Group Manager (GM) selects two additive group $G_1$, $G_2$ and one multiplicative group $G_3$ with order q. GM shares a generator P for $G_2$ publicly. A polynomial with degree t-1 is chosen. First coefficient of the polynomial will be the master secret s. Q=s.P is computed and shared publicly. For each user, one public key $x_i$ and one private key f($x_i$) are chosen and shared with related users secretly.Users participating the authentication phase aggre on a random point $R_v$ on $G_1$ in authentication phase. Then, each user computes $c_i$=f($x_i$).${\overset{m}{\underset{r=i, r\neq i}{{\displaystyle\prod}}}(-x_r/(x_i-x_r))}$ and releases $c_i$.$R_v$. After all users release the $c_i$.$R_v$, each user computes ${\overset{m}{\underset{i=1}{{\displaystyle\sum}}}c_i.R_v}$ and verifies e(${\overset{m}{\underset{i=1}{{\displaystyle\sum}}}c_i.R_v}$ ,P) ${\stackrel{?}{=}}$  e($R_v$,Q) holds. The algorithm provides security for group authentication except Node Comprimise and DOS attack but it is resource consuming for users.

\section{Proposed Method}
Our proposal for group authentication is based on ECC and Shamir Secret Sharing. There are two phases which are initialisation and authentication phases.
\\\\
In initialisation phase:
\\
Step 1. GM selects an additive group G and a generator P for G.
\\
Step 2. GM selects a bilinear map e: G x G and an Encryption(E)/Decryption(D) algorithm.
\\
Step 3. A polynomial with degree t-1 is chosen by GM and first coefficient is determined as secret s.
\\
Step 4. GM selects one public key $x_i$ and one private key f($x_i$) for each user in the group.(U={$U_1$,$U_2$,...,$U_n$}).
\\
Step 5. GM computes Q=s.P.
\\
Step 6. GM shares P,Q,e,E/D publicly and f($x_i$) secretly.
\\\\
In authentication phase:
\\
Step 1. Each user computes f($x_i$).P and sends  f($x_i$).P$\|ID_i$ to GM and other users to be used for key generating.
\\
Step 2. GM computes $c_i$=f($x_i$).P.${\overset{m}{\underset{r=i, r\neq i}{{\displaystyle\prod}}}(-x_r/(x_i-x_r))}$ for each user.
\\
Step 3. GM verifies ${\overset{m}{\underset{i=1}{{\displaystyle\sum}}}c_i}$  ${\stackrel{?}{=}}$  Q holds.
\\
Step 4. If it holds, authentication is done. Otherwise; GM will repeat the proces from the initialisation phase.
\\\\
After authentication is done, users will communicate with each other by using symmetric key encryption. Shared key for symmetric key encryption will be calculated by senders and receivers.
\\
Key= e($y_i.y_j.P$,Q),i=sender,j=receiver. Sender will use its own private key and value sent by receiver and Q. Receiver will compute same key by using its own private key, value sent by sender and Q.
\\\\
In our proposal, we used the same group authentication schema (t,m,n) in Harn algorithm. There are n users in the group and m users want to authenticate each other. t is the threshold for the algorithm (t$<m<n$). n should be greater than m and secret can be obtained by the participation of m users or more users till n users. 
\\\\
GM always knows that m user partipated the authentication and m+1 value is not used yet. So in the hand-over phase:
\\
Step 1. $GM_2$ select m+1 value which is not used in authentication and computes f($x_m+1$).P and sends other GMs secretly.
\\
Step 2. If any user from other groups want to particapate Group 2, user gets the f($x_m+1$).P from its GM.
\\
Step 3. New user sends the f($x_m+1$).P to $GM_2$.
\\
Step 4. $GM_2$ computes $c_m$=f($x_m$).P.${\overset{m+1}{\underset{r=i, r\neq i}{{\displaystyle\prod}}}(-x_r/(x_i-x_r))}$.
\\
Step 5.  $GM_2$  verifies ${\overset{m+1}{\underset{i=1}{{\displaystyle\sum}}}c_i}$  ${\stackrel{?}{=}}$  s.P holds.
\\
Step 6. If it holds, the hand-over porcess is valid and$GM_2$ updates other group managers.
\\
Step 7. Even if it holds or not, $GM_2$  selects m+2 which is not used for authentication and computes f($x_m+2$).P.
\\
Step 8. $GM_2$   sends f($x_m+2$).P to other GMs secretly.
\\\\
We compared the computational costs of Harn, Chien and our algorithm. According to Chien, their algorithm has (7m+6785)$T_{mul,q}$ computational cost and Harn algorithm has (45m+1418)$T_{mul,q}$ computational cost (m denote the number of users, $T_{mul,q}$ denote that for one multiplication in the field q). IoT users have to compute only one Eliptic Curve point multiplication in our proposal and it costs 1189$T_{mul,q}$ [12].  
\begin{figure}[h!]
  \includegraphics[width=\linewidth]{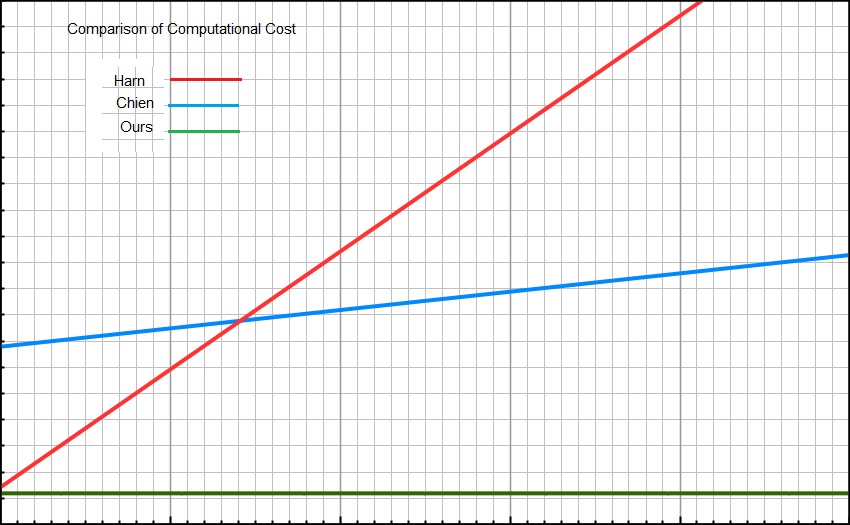}
  \caption{Computational costs of Harn, Chien Algorithms and Our Proposal.}
  \label{fig:boat1}
\end{figure}
\section{Conclusion}
The study proposed a novel method for authentication and hand-over process for IoT groups.
Many-to-many authentication is used for Group Authentication by several studies but resource-constrained IoT nodes should compute more than their capacity. 
IoT nodes sould only compute one multiplication in the proposed method.

\end{document}